\documentclass[amsmath,amssymb]{revtex4}
\usepackage{graphicx}
\usepackage{dcolumn}
\usepackage{bm}

\newcommand{\gsim}{\raisebox{0.2ex}{$\ > \kern -1.05em%
        \raisebox{-1.1ex}{$\sim$}\ $}}  
\begin{document}

\title{External Source Method for 
Kubo-Transformed Quantum Correlation Functions
}

\author{Atsushi Horikoshi} 
\email{horikosi@tcu.ac.jp}
\affiliation{
Department of Natural Sciences, Faculty of Knowledge Engineering, Tokyo City University, 
Tamazutsumi, Setagaya-ku, Tokyo 158-8557, Japan
}

\begin{abstract}
We revisit the external source method 
for Kubo-transformed quantum correlation functions
recently proposed by Krishna and Voth.
We derive an exact formula and show that 
the Krishna-Voth formula can be derived 
as an approximation of our formula. 
Some properties of this approximation are clarified through 
a model calculation of the position autocorrelation function 
for a one-dimensional harmonic oscillator.
A key observation is that 
the Krishna-Voth correlation function has a term 
which behaves as the secular term in perturbation theory.
\end{abstract}

\maketitle

\section{Introduction} 
\hspace*{\parindent}
The Kubo-transformed quantum correlation function $C^{\rm K}_{AB}(t)$
is an important quantity to characterize quantum dynamical effects 
of operators $\hat{A}$ and $\hat{B}$
and plays a central role in the linear response theory \cite{KuboTodaHashitsume}.
However, it is not easy to calculate $C^{\rm K}_{AB}(t)$, 
because most of the numerical techniques 
for quantum many-body systems \cite{BerneThirumalai,Ceperley}
are based on the imaginary time path integral formalism \cite{FeynmanHibbs}
and we cannot directly apply them to the calculation of $C^{\rm K}_{AB}(t)$.
\par
Recently, a new quantum dynamics method to calculate $C^{\rm K}_{AB}(t)$ 
has been proposed by Krishna and Voth \cite{KrishnaVoth}.
This is an extended quantum dynamics perturbed 
by external sources \cite{Schwinger,Kleinert}  
and is applicable to the case that 
the operators $\hat{A}$ and $\hat{B}$ depend on 
the position operator $\hat{q}$ only.
The calculation procedure is summarized as follows:
(1) Introduce two external sources $\mu$ and $\nu$ that couple to
the operators $\hat{A}$ and $\hat{F}$, respectively; 
here, $\hat{F}$ is the operator that satisfies
$\hat{B}=\frac{\partial}{\partial q}\hat{F}$.
(2) Calculate the expectation value of the momentum operator 
$\hat{p}$, $p_{\mu\nu}(t)={\rm Tr}(\hat{\rho}_{\mu}\hat{p}_{\nu}(t))$,
where the equilibrium density operator $\hat{\rho}_{\mu}$
is given by the Hamiltonian $\hat{H}+{\mu}\hat{A}$
and the time evolution of $\hat{p}$ is 
driven by the Hamiltonian $\hat{H}+{\nu}\hat{F}$.
(3) Differentiate $p_{\mu\nu}(t)$ with respect to $\mu$, $\nu$, and $t$;
we obtain the Kubo-transformed quantum correlation function using the 
Krishna-Voth formula 
$C^{\rm K}_{AB}(t)= 
\frac{1}{\beta}\frac{\partial^{3}}{\partial\mu\partial\nu\partial t}
~p_{\mu\nu}(t)
+\langle\hat{A}\rangle_{\beta} \langle\hat{B}\rangle_{\beta}$.
Their method is quite promising, 
because $p_{\mu\nu}(t)$ is a nonequilibrium expectation value
that can be calculated 
by means of various quantum dynamics 
methods \cite{JangVoth,Jang,Miller,Tanimura,CoffeyKalmykovTitovCleary}.
They have applied this new method to a quantum anharmonic oscillator
and shown that it works well for the calculation of 
nonlinear correlation functions \cite{KrishnaVoth}.
It is still unclear, however, whether their method can be applied to
general quantum systems,  
because basic properties of the Krishna-Voth formula 
have not been clarified.
\par
In this work, we refine the formulation of 
Krishna and Voth's external source method,
and derive a new exact formula 
for Kubo-transformed quantum correlation functions.
We clarify the approximation implicitly used by Krishna and Voth
and show that the approximation is valid in the short time limit.
As an example, we show the calculations of the position autocorrelation function 
for a harmonic oscillator and discuss some properties 
of this method.

\section{External Source Method}
\hspace*{\parindent}
In this section, we present a refined formulation of 
Krishna and Voth's external source method \cite{KrishnaVoth}.
For simplicity, we consider a one-dimensional quantum system 
described by Hamiltonian 
\begin{eqnarray}
\hat{H}=\frac{1}{2m}\hat{p}^{2}+V(\hat{q}).
\label{1}
\end{eqnarray}
The equilibrium density operator of this system is given by
$\hat{\rho}_{\beta}=e^{-\beta \hat{H}}/Z_{\beta}$,
where $\beta=1/(k_{B}T)$ is the inverse temperature
and 
$Z_{\beta}={\rm Tr}(e^{-\beta\hat{H}})$ is the 
quantum partition function.
The thermal expectation value of an operator $\hat{A}$ 
is defined by
$\langle\hat{A}\rangle_{\beta}={\rm Tr}(\hat{\rho}_{\beta}\hat{A})$.
We consider a Kubo-transformed quantum correlation function 
for position dependent operators $\hat{A}=A(\hat{q})$ and $\hat{B}=B(\hat{q})$,
\begin{eqnarray}
C^{\rm K}_{AB}(t)=
\langle \hat{A}^{\rm K}_{\beta}(0)\hat{B}(t)\rangle_{\beta},
\label{2}
\end{eqnarray}
where $\hat{A}^{\rm K}_{\beta}$ is the Kubo-transformed operator,
\begin{eqnarray}
\hat{A}^{\rm K}_{\beta}=\frac{1}{\beta}\int^{\beta}_{0}\!d\lambda
~e^{\lambda\hat{H}}\hat{A}~e^{-\lambda\hat{H}}.
\label{3}
\end{eqnarray}
\subsection{Two external sources and two perturbed Hamiltonians}
\hspace*{\parindent}
First, we introduce a constant external source $\mu$ which couples to
the operator $\hat{A}$. The perturbed Hamiltonian is defined by
\begin{eqnarray}
\hat{H}_{\mu}=\hat{H}+\mu\hat{A}.
\label{4}
\end{eqnarray}
Assuming small $\mu$, we expand the Boltzmann operator for 
$\hat{H}_{\mu}$,
\begin{eqnarray}
e^{-{\beta}\hat{H}_{\mu}}=e^{-\beta \hat{H}}
\left(
\hat{1}-\mu\int^{\beta}_{0}\!d\lambda
~e^{\lambda\hat{H}}\hat{A}~e^{-\lambda\hat{H}}
+O(\mu^{2})\right).
\label{5}
\end{eqnarray}
This leads to following identities:
\begin{eqnarray}
\left.\frac{\partial}{\partial\mu}
e^{-\beta\hat{H}_{\mu}}\right|_{\mu=0}&=&
-\beta e^{-\beta\hat{H}}\hat{A}^{\rm K}_{\beta},\label{6}\\
\left.\frac{\partial}{\partial\mu}
Z_{\mu}\right|_{\mu=0}&=&
-\beta Z_{\beta}\langle\hat{A}\rangle_{\beta},\label{7}\\
\left.\frac{\partial}{\partial\mu}
\hat{\rho}_{\mu}\right|_{\mu=0}&=&
-\beta \hat{\rho}_{\beta}
(\hat{A}^{\rm K}_{\beta}-\langle\hat{A}\rangle_{\beta}).\label{8}
\end{eqnarray}
Here, $Z_{\mu}={\rm Tr}(e^{-\beta\hat{H}_{\mu}})$
is the quantum partition function and 
$\hat{\rho}_{\mu}=e^{-\beta \hat{H}_{\mu}}/Z_{\mu}$
is the equilibrium density operator of the perturbed system.
\\
\hspace*{\parindent}
Next, we consider a position dependent operator $\hat{F}=F(\hat{q})$ 
that satisfies
\begin{eqnarray}
B(\hat{q})=\frac{\partial}{\partial q}F(\hat{q}),
\label{9}
\end{eqnarray}
and introduce another external source $\nu$ that couples to
the operator $\hat{F}$. 
Another perturbed Hamiltonian is defined by
\begin{eqnarray}
\hat{H}_{\nu}=\hat{H}+\nu\hat{F}.
\label{10}
\end{eqnarray}
Then, we consider a time-dependent momentum operator
$\hat{p}_{\nu}(t)$, the time evolution of which is given by $\hat{H}_{\nu}$,
\begin{eqnarray}
\hat{p}_{\nu}(t)=e^{i\hat{H}_{\nu}t/\hbar}~\hat{p}~e^{-i\hat{H}_{\nu}t/\hbar}.
\label{11}
\end{eqnarray}
Using perturbative expansions of the time evolution operators with respect to $\nu$, 
\begin{eqnarray}
e^{\pm i\hat{H}_{\nu}t/\hbar}=e^{\pm i\hat{H}t/\hbar}
\left(
\hat{1}\pm\nu\frac{i}{\hbar}\int^{t}_{0}\!ds
~e^{\mp i\hat{H}s/\hbar}\hat{F}~e^{\pm i\hat{H}s/\hbar}+O(\nu^{2})
\right),
\label{12}
\end{eqnarray}
and using an identity
\begin{eqnarray}
\frac{i}{\hbar}\left[\hat{p},F(\hat{q})\right]=\frac{\partial}{\partial q}F(\hat{q}),
\label{13}
\end{eqnarray}
where $[~,~]$ is the commutator
$[\hat{A},\hat{B}]=\hat{A}\hat{B}-\hat{B}\hat{A}$,
we obtain the following identity
\begin{eqnarray}
\left.\frac{\partial^{2}}{\partial\nu\partial t}\hat{p}_{\nu}(t)\right|_{\nu=0}
=-\hat{B}(t)+\hat{D}(t).
\label{15}
\end{eqnarray}
Here, $\hat{D}(t)$ is an operator,
\begin{eqnarray}
\hat{D}(t)=-\frac{i}{\hbar}
\left[
\int^{t}_{0}\!ds\hat{F}(s),~\frac{\partial \hat{V}}{\partial q}(t)
\right],
\label{15a}
\end{eqnarray}
which is omitted in the Krishna-Voth formalism \cite{KrishnaVoth}.
Note that $\hat{D}(t)$ vanishes at $t=0$, that is, $\hat{D}(0)=0$.
\subsection{Exact formula 
for Kubo-transformed quantum correlation functions}
\hspace*{\parindent}
Let us rewrite the Kubo-transformed quantum correlation function
$C^{\rm K}_{AB}(t)$ (Eq. (\ref{2})) by means of 
two external sources ($\mu$, $\nu$) and corresponding
perturbed Hamiltonians ($\hat{H}_{\mu}$, $\hat{H}_{\nu}$).
Consider a time-dependent expectation value of the momentum operator $\hat{p}$,
\begin{eqnarray}
p_{\mu\nu}(t)=
{\rm Tr}(\hat{\rho}_{\mu}\hat{p}_{\nu}(t)).
\label{16}
\end{eqnarray}
This is a nonequilibrium expectation value of $\hat{p}$,
because the initial distribution is given by one Hamiltonian 
whereas the time evolution is driven by another. 
The derivative of $p_{\mu\nu}(t)$
with respect to $\mu$, $\nu$, and $t$ 
at $\mu=0$ and $\nu=0$
can be written 
using Eq. (\ref{15}) as 
\begin{eqnarray}
\left.\frac{\partial^{3}}{\partial\mu\partial\nu\partial t}
~p_{\mu\nu}(t)\right|_{\mu,\nu=0}=
-{\rm Tr}\left(\left.\frac{\partial \hat{\rho}_{\mu}}{\partial \mu}\right|_{\mu=0}
\!\!\!\!\!\!\!\hat{B}(t)\right)
+{\rm Tr}\left(\left.\frac{\partial \hat{\rho}_{\mu}}{\partial \mu}\right|_{\mu=0}
\!\!\!\!\!\!\!\hat{D}(t)\right).
\label{17}
\end{eqnarray} 
Using Eq. (\ref{8}),
we obtain a new exact formula for 
Kubo-transformed quantum correlation functions, 
\begin{eqnarray}
C^{\rm K}_{AB}(t)=
\frac{1}{\beta}\left.\frac{\partial^{3}}{\partial\mu\partial\nu\partial t}
~p_{\mu\nu}(t)\right|_{\mu,\nu=0}
+\langle\hat{A}\rangle_{\beta} \langle\hat{B}\rangle_{\beta}
+C^{\rm K}_{AD}(t)-\langle\hat{A}\rangle_{\beta} \langle\hat{D}(t)\rangle_{\beta},
\label{18}
\end{eqnarray}
where $C^{\rm K}_{AD}(t)=
\langle \hat{A}^{\rm K}_{\beta}(0)\hat{D}(t)\rangle_{\beta}$.
This formula allows us to calculate $C^{\rm K}_{AB}(t)$
by pursuing the dynamics of the momentum expectation value $p_{\mu\nu}(t)$ 
and the operator $\hat{D}(t)$. 
\\
\hspace*{\parindent}
It should be noted here that 
there is a restriction on the operators $\hat{A}$ and $\hat{B}$. 
The operators we can treat in this external source method 
are limited depending on the Hamiltonian of the original system. 
This is because the spectra of the perturbed Hamiltonians 
$\hat{H}_{\mu}$ and $\hat{H}_{\nu}$ should be bounded from below
for the stability of the perturbed systems.

\subsection{Krishna-Voth approximation}
\hspace*{\parindent}
Although Eq. (\ref{18}) is exact, this formula is not 
suited for practical uses,
because the calculations of 
$C^{\rm K}_{AD}(t)$ and $\langle\hat{D}(t)\rangle_{\beta}$
are in general more demanding than the calculation of $C^{\rm K}_{AB}(t)$. 
For this reason, it might be useful to give an approximate expression 
of Eq. (\ref{18}). 
As a simple approximation, 
we neglect the contribution of the operator $\hat{D}$
in Eq. (\ref{15})
\begin{eqnarray}
\hat{D}(t)=0, 
\label{19a}
\end{eqnarray}
and obtain a simple formula
\begin{eqnarray}
C^{\rm K}_{AB}(t)\simeq
C^{\rm KV}_{AB}(t)=
\frac{1}{\beta}\left.\frac{\partial^{3}}{\partial\mu\partial\nu\partial t}
~p_{\mu\nu}(t)\right|_{\mu,\nu=0}
+\langle\hat{A}\rangle_{\beta} \langle\hat{B}\rangle_{\beta}.
\label{19}
\end{eqnarray}
This is identical to the formula given by Krishna and Voth \cite{KrishnaVoth}.
Therefore, we refer to the approximation (Eq. (\ref{19a})) as 
the Krishna-Voth approximation
and refer to $C^{\rm KV}_{AB}(t)$ (Eq. (\ref{19}))
as the Krishna-Voth correlation function. 
Because of $\hat{D}(0)=0$ for any system,
the Krishna-Voth approximation is exact in the short time limit.
\section{Results and Discussion}
\subsection{Application to a quantum harmonic oscillator}
\hspace*{\parindent}
Let us apply the external source method (Eqs. (\ref{18}) and (\ref{19}))) 
to the calculations of 
Kubo-transformed autocorrelation functions 
$C^{\rm K}_{AA}(t)=\langle \hat{A}^{\rm K}_{\beta}(0)\hat{A}(t)\rangle_{\beta}$
for a one-dimensional harmonic oscillator, the potential
of which is given by
\begin{eqnarray}
V(\hat{q})=\frac{m\omega^{2}}{2}\hat{q}^{2}.
\label{20}
\end{eqnarray}
In the harmonic system, 
if the operator $\hat{A}$ is nonlinear in position, 
$\hat{A}=\hat{q}^{n} (n\geq 2)$,
either $\hat{H}_{\mu}$ or $\hat{H}_{\nu}$ always gives
energy spectra unbounded from below.
Therefore, we treat the linear operator, $\hat{A}=\hat{q}$,
to illustrate the method. 
The exact expression of 
$C^{\rm K}_{qq}(t)=\langle \hat{q}^{\rm K}_{\beta}(0)\hat{q}(t)\rangle_{\beta}$
can be written as 
\begin{eqnarray}
C^{\rm K}_{qq}(t)=\frac{1}{\beta m\omega^{2}}\cos\omega t.
\label{21}
\end{eqnarray}
In this case, the perturbations are introduced as  
$\mu\hat{A}=\mu\hat{q}$ and $\nu\hat{F}=\frac{\nu}{2}\hat{q}^{2}$,
where the external source $\nu$ is assumed to be 
$\nu>-m\omega^{2}$ for the stability of the perturbed system.
The time dependent momentum operator (Eq. (\ref{11})) and 
its nonequilibrium expectation value (Eq. (\ref{16})) can be calculated analytically,
\begin{eqnarray}
\hat{p}_{\nu}(t)&=&-\hat{q}(0)m\Omega\sin\Omega t +\hat{p}(0)\cos\Omega t,
\label{22}\\
p_{\mu\nu}(t)&=&\frac{\mu\Omega}{\omega^{2}}\sin\Omega t,
\label{23}
\end{eqnarray}
where $\Omega=\sqrt{\omega^{2}+\nu/m}$
is the modified frequency.
The derivative of $p_{\mu\nu}(t)$
with respect to $\mu$, $\nu$, and $t$
is evaluated at $\mu=0$ and $\nu=0$ as
\begin{eqnarray}
\left.\frac{\partial^{3}}{\partial\mu\partial\nu\partial t}
~p_{\mu\nu}(t)\right|_{\mu,\nu=0}=
\frac{1}{m\omega^{2}}\cos\omega t-\frac{t}{2m\omega}\sin\omega t.
\label{24}
\end{eqnarray}
On the other hand, the operator $\hat{D}(t)$ is given by
\begin{eqnarray}
\hat{D}(t)=\hat{q}(0)\frac{\omega t\sin\omega t}{2}+
\hat{p}(0)\frac{\sin\omega t-\omega t\cos\omega t}{2m\omega},
\label{25}
\end{eqnarray}
and another Kubo-transformed quantum
correlation function $C^{\rm K}_{AD}(t)$ is obtained as 
\begin{eqnarray}
C^{\rm K}_{AD}(t)=C^{\rm K}_{qD}(t)=
\frac{t}{2\beta m\omega}\sin\omega t.
\label{26}
\end{eqnarray}
We also obtain $\langle\hat{D}(t)\rangle_{\beta}=0$ and 
$\langle\hat{A}\rangle_{\beta}=\langle\hat{B}\rangle_{\beta}
=\langle\hat{q}\rangle_{\beta}=0$.
Although Eqs. (\ref{24}) and (\ref{26}) contain unfavorable terms that
diverge in the long time limit, these terms are cancelled out in Eq. (\ref{18})
to reproduce the exact result (Eq. (\ref{21})).
Finally, we obtain the Krishna-Voth correlation function (Eq. (\ref{19})),
\begin{eqnarray}
C^{\rm KV}_{qq}(t)&=&
\frac{1}{\beta m\omega^{2}}\cos\omega t-\frac{t}{2\beta m\omega}\sin\omega t
\label{27}\\
&=&
\frac{1}{\beta m\omega^{2}}\sqrt{1+\frac{\omega^{2}t^{2}}{4}}
\cos(\omega t +\alpha),
\label{28}
\end{eqnarray}
where $\alpha=\tan^{-1} (\omega t/2)$.
\subsection{Secular term}
\begin{figure}
\begin{tabular}{c}
\includegraphics[width=150mm]{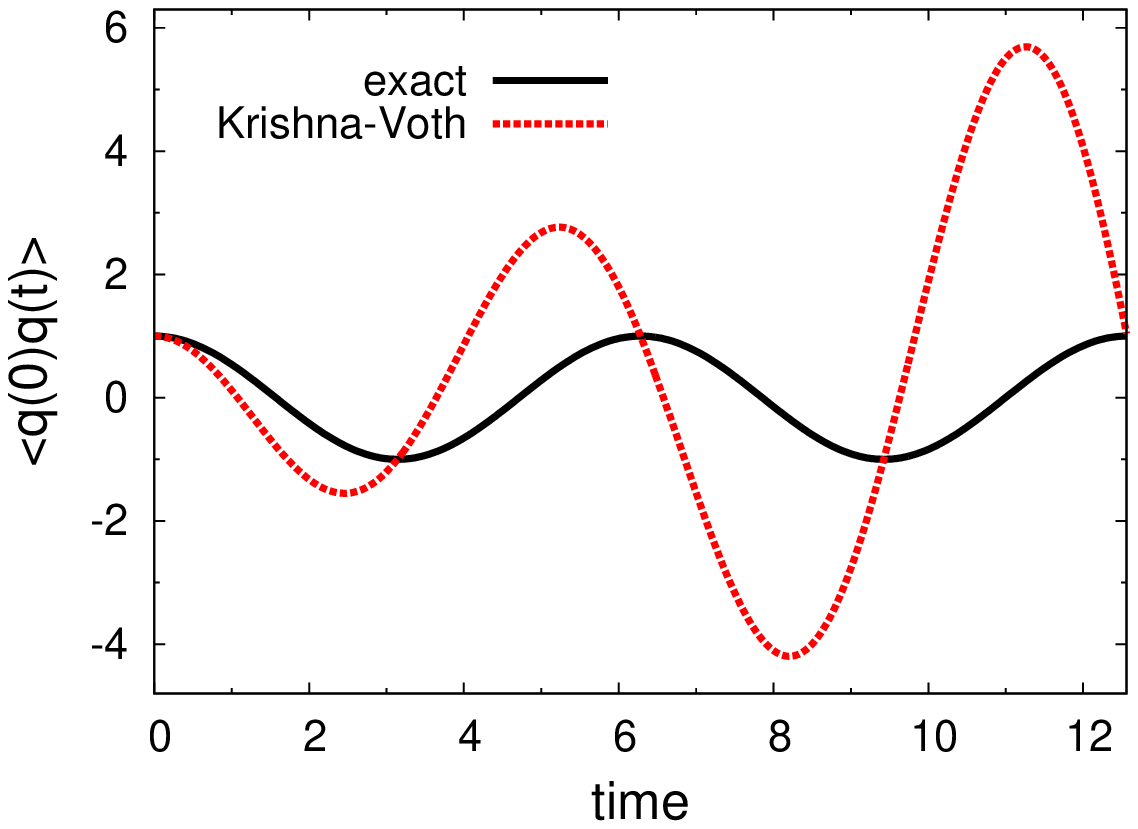}
\end{tabular}
\vspace{0mm}
\caption{Kubo-transformed position autocorrelation functions for 
the harmonic oscillator (Eq. (\ref{20})).
Solid line: the exact correlation function 
$C^{\rm K}_{qq}(t)$ (Eq. (\ref{21})).
Dashed line: the Krishna-Voth correlation function 
$C^{\rm KV}_{qq}(t)$ (Eq. (\ref{28})).
}
\label{fig1}
\end{figure}

\hspace*{\parindent}
Figure \ref{fig1} shows the plot of 
the exact Kubo-transformed quantum correlation function (Eq. (\ref{21}))
and the Krishna-Voth correlation function (Eq. (\ref{28}))
with the parameters $\hbar=k_{B}=m=\omega=\beta=1$.
The Krishna-Voth correlation function coincides with 
the exact correlation function at $t=0$. 
This is because the Krishna-Voth approximation (Eq. (\ref{19a}))
is exact in the short time limit.
However, as the time $t$ increases,  
the Krishna-Voth correlation function deviates from the exact one
and its amplitude grows with time. 
In the long time limit $t \to \infty$, 
the phase shift $\alpha$ converges to $\pi/2$,
whereas the amplitude enhancement factor
$\sqrt{1+\omega^{2}t^{2}/4}$
diverges.
Such unbounded growth of amplitude is often observed  
in perturbation theory \cite{BenderOrszag}.
The second term of Eq. (\ref{27}), 
which is proportional to $t\sin\omega t$,
corresponds to the secular term 
in perturbation theory \cite{BenderOrszag}.
Therefore, for the calculation of the position autocorrelation function for 
the harmonic oscillator,
the Krishna-Voth approximation (Eq. (\ref{19a}))
works only in the short time region.
The discrepancy between $C^{\rm K}_{qq}(t)$ and $C^{\rm KV}_{qq}(t)$
can be improved by neither 
increasing nor decreasing the temperature $T$. 
This is because in this case, 
the ratio of $C^{\rm KV}_{qq}(t)$ to $C^{\rm K}_{qq}(t)$,
\begin{eqnarray}
\frac{C^{\rm KV}_{qq}(t)}{C^{\rm K}_{qq}(t)}=
1-\frac{\omega t}{2}\tan\omega t,
\label{29}
\end{eqnarray}
is independent of the temperature.
Therefore, at any temperature,
we obtain the result similar in shape as 
Figure \ref{fig1}. 
\\
\hspace*{\parindent}
Here, we give a brief discussion on
the origin of the secular behavior of $C^{\rm KV}_{AB}(t)$ (Eq. (\ref{19})).
Let us start with an eigenstate representation of $p_{\mu\nu}(t)$,
\begin{eqnarray}
p_{\mu\nu}(t)
=\sum_{n}\sum_{m}
e^{-i(E_{n}(\nu)-E_{m}(\nu))t/\hbar}
\langle \phi_{n}(\nu)|\hat{\rho}_{\mu}|\phi_{m}(\nu)\rangle
\langle \phi_{m}(\nu)|\hat{p}|\phi_{n}(\nu)\rangle,
\label{30}
\end{eqnarray}
where $|\phi_{n}(\nu)\rangle$ and $E_{n}({\nu})$
are the eigenstates and eigenvalues of the perturbed 
Hamiltonian $\hat{H}_{\nu}$ (Eq. (\ref{10})). 
The derivative of $p_{\mu\nu}(t)$ with respect to $\mu$, $\nu$, and $t$
contains terms proportional to
$t~e^{-i(E_{n}({\nu})-E_{m}({\nu}))t/\hbar}$.
How these terms contribute to $C^{\rm KV}_{AB}(t)$
depends on the character of the system
and the functional forms of operators $\hat{A}$ and $\hat{B}$.
As we have shown in this section, 
for the one-dimensional harmonic oscillator,
the secular term becomes dominant  
in $C^{\rm KV}_{qq}(t)$
as the time $t$ increases.
On the other hand,
Krishna and Voth have applied the method 
to a one-dimensional anharmonic oscillator 
$V(\hat{q})=\frac{1}{2}\hat{q}^{2}+
\frac{1}{10}\hat{q}^{3}+
\frac{1}{100}\hat{q}^{4}$
and computed nonlinear correlation functions
$C^{\rm KV}_{q^{2}q^{2}}(t)$ and 
$C^{\rm KV}_{q^{3}q^{3}}(t)$ \cite{KrishnaVoth}.
Their numerical results show no sign of diverging behavior
and agree reasonably 
well with the exact results in a wide time range. 
Their work suggests that 
the contribution of the terms proportional to
$t~e^{-i(E_{n}({\nu})-E_{m}({\nu}))t/\hbar}$
could be suppressed by non-linearity of 
the potential  and/or the operators.  

\section{Conclusions}
\hspace*{\parindent}
In this work, we have reformulated 
the external source method proposed by Krishna and Voth \cite{KrishnaVoth}
and derived an exact formula for 
Kubo-transformed quantum correlation functions (Eq. (\ref{18})).
An approximation (Eq. (\ref{19a})), which is exact at $t=0$,
have been proposed and identified 
as the approximation implicitly used by Krishna and Voth \cite{KrishnaVoth}.
We have carried out analytical calculations of 
the position autocorrelation functions 
for the one-dimensional harmonic oscillator (Eq. (\ref{20}))
and found that the Krishna-Voth correlation function (Eq. (\ref{27}))
has a term that behaves as the secular term in perturbation theory.
\\
\hspace*{\parindent}
It will be important in future work
to clarify the condition that
the Krishna-Voth approximation (Eq. (\ref{19a}))
remains valid beyond the short time region.
More detailed analyses based on 
the eigenstate representation (Eq. (\ref{30}))
will be needed.
The test calculations performed by Krishna and Voth \cite{KrishnaVoth}
suggest that the secular behavior of $C^{\rm KV}_{AB}(t)$
could be suppressed by 
nonlinearity of the potential $V(\hat{q})$ and/or
nonlinearity of the operators $\hat{A}$ and $\hat{B}$.
In addition to these nonlinear effects,
symmetry and multi-dimensionality of the system might be significant.
This is because, in general, 
symmetry of the system could give some constraints on 
the behavior of correlation functions,
and many-body effects could cause dephasing, that is,
suppression of the oscillating behavior of correlation functions.
It will also be interesting to 
examine the possibility to
get rid of the secular terms
by means of some prescriptions 
like the renormalization group method
for ordinary differential equations \cite{ChenGoldenfeldOono}.
\\
\hspace*{\parindent}
In the external source method,
the types of operators $\hat{A}$ and $\hat{B}$ are 
limited by a stability condition:
the spectra of the perturbed Hamiltonians 
$\hat{H}_{\mu}$ and $\hat{H}_{\nu}$
should be bounded from below to ensure
the stability of the perturbed systems.
However, in practice, the condition could be relaxed,
because the method is formulated 
by means of the perturbative expansion
with respect to infinitesimal external sources $\mu$ and $\nu$,
and, therefore, the equilibrium distribution and the time evolution 
of the unstable perturbed systems could be effectively stable. 
In analytical calculations,
we could make a rigorous discussion
on the effective stability of unstable perturbed systems,
because we can take the limit $\mu, \nu \to 0$ analytically.
On the other hand,
in numerical calculations,
more careful handling of the effectively stable systems
would be required,
because in numerical treatments,
the external sources $\mu$ and $\nu$
are chosen to be small but finite
to evaluate the derivatives of $p_{\mu\nu}$ by 
the finite difference approximation, for example,
$\frac{\partial{p_{\nu}}}{\partial\nu}\simeq
\frac{p_{\nu}-p_{-\nu}}{2\nu}$ \cite{KrishnaVoth}. 
Therefore, some practical prescriptions to handle such perturbed systems
should be established.
\\
\hspace*{\parindent}
It is also important to develop efficient algorithms 
to calculate the time-dependent momentum expectation value
$p_{\mu\nu}(t)$ (Eq. (\ref{16})).
By means of 
the nonequilibrium path integral centroid dynamics \cite{JangVoth},
$p_{\mu\nu}(t)$ for the position autocorrelation functions 
for the one-dimensional harmonic oscillator (Eq. (\ref{23}))
can be calculated exactly \cite{Horikoshi}.
However, the numerical implementation of
the nonequilibrium path integral centroid dynamics
could be much more demanding than 
the usual path integral molecular dynamics \cite{BerneThirumalai}.
Therefore, 
other nonequilibrium quantum dynamics 
methods \cite{KrishnaVoth,Jang,Miller,Tanimura,CoffeyKalmykovTitovCleary} 
would be better suited for further applications of 
the external source method 
to more realistic systems.



\newpage
\end{document}